\documentclass[
  aps,
  prb,
  twocolumn,
  superscriptaddress,
  showkeys,
  longbibliography
]{revtex4-2}

\usepackage{amsmath,amssymb,bm}

\usepackage{graphicx, tikz, subfigure}
\usepackage{subcaption} % subfigure environment
\usepackage[percent]{overpic} % подписи поверх рисунка без TikZ

\usepackage[colorlinks=true,
            linkcolor=blue,
            citecolor=blue,
            urlcolor=blue]{hyperref}
\usepackage{orcidlink}

\hypersetup{
  pdftitle={Sub-domain structure in a single crystal of the magnetic topological insulator MnSb2Te4},
  pdfauthor={V. A. Tyutvinov; M. S. Sidelnikov; N. A. Abdullayev; Z. S. Aliev; I. R. Amiraslanov; N. T. Mamedov; V. N. Zverev; L. Ya. Vinnikov},
  pdfsubject={High-resolution Bitter decoration reveals two-scale magnetic domain structure in MnSb2Te4 (Tc ~45 K) indicating weak surface-bulk coupling due to MnSb site mixing and surface symmetry breaking},
  pdfkeywords={MnSb2Te4, MnBi2Te4 family, magnetic topological insulator, magnetic domains, sub-domain structure, two-scale domain structure, Bitter decoration technique, antisite defects, surface-bulk magnetic coupling, defect-induced ferromagnetism, symmetry breaking}
}

\usepackage{physics}
\usepackage{braket}

\usepackage{ragged2e}

\begin{document}

\title{Sub-domain structure in a single crystal of the magnetic topological insulator MnSb$_2$Te$_4$}

\author{V.\,A.~Tyutvinov\,\orcidlink{0009-0006-3652-8175}}
\email{tiutvinov.va@phystech.edu}
% Если надо указать рядом с почтой где опубликована статья
%\thanks{\\Accepted for publication in JETP Letters}
\affiliation{Institute of Solid State Physics, Russian Academy of Sciences, 142432 Chernogolovka, Russia}
\affiliation{Moscow Institute of Physics and Technology, 141700 Dolgoprudny, Russia}

\author{M.\,S.~Sidelnikov\,\orcidlink{0000-0003-2215-6703}}
\affiliation{Institute of Solid State Physics, Russian Academy of Sciences, 142432 Chernogolovka, Russia}

\author{N.\,A.~Abdullayev\,\orcidlink{0000-0002-0276-5985}}
\author{Z.\,S.~Aliev\,\orcidlink{0000-0001-5724-4637}}
\author{I.\,R.~Amiraslanov\,\orcidlink{0000-0001-7975-614X}}
\author{N.\,T.~Mamedov\,\orcidlink{0009-0009-4302-1913}}
\affiliation{Institute of Physics, Ministry of Science and Education, AZ 1073 Baku, Azerbaijan}
\affiliation{Baku State University, AZ 1148 Baku, Azerbaijan}

\author{V.\,N.~Zverev\,\orcidlink{0000-0002-2857-7587}}
\author{L.\,Ya.~Vinnikov\,\orcidlink{0000-0001-5601-3925}}
\affiliation{Institute of Solid State Physics, Russian Academy of Sciences, 142432 Chernogolovka, Russia}

\begin{abstract}
The domain structure of a MnSb$_2$Te$_4$ single crystal with a Curie temperature $T_C \approx 45$ K was studied using the high-resolution Bitter decoration technique. Magnetotransport measurements confirm a soft ferromagnetic ordering with a coercive field of $ \sim 100$ Oe. We revealed the formation of a hierarchical domain structure characterized by two distinct spatial scales. These results indicate the existence of two magnetically weakly coupled subsystems --- surface and bulk. The observed sub-domain structure can be attributed to the formation of a ferromagnetic well due to an inhomogeneous distribution of $\mathrm{Mn_{Sb}}$ antisite defects, with an additional contribution from symmetry breaking in the near-surface layer.
\end{abstract}


\maketitle

\section{Introduction}

\label{introduction}

The discovery of intrinsic magnetic topological insulators, such as $\mathrm{MnBi_2Te_4}$ (MBT), has stimulated intensive research on the quantum anomalous Hall effect (QAHE) \cite{otrokov2019firstAFMMBT, deng2020QAHE, tokura2019review, bernevig2022review, vyazovskaya2025review}, the axion insulator state \cite{liu2020axionchern}, and Weyl semimetals \cite{li2021weyl}. The prospect of realizing the QAHE in the MBT system without an external magnetic field opens up opportunities for creating energy-efficient spintronic devices. However, practical implementations are hindered by fundamental limitations: strong $n$-type bulk conductivity in MBT \cite{chen2019n-type}, which complicates the observation of chiral edge states \cite{zhu2025edgecurrentproblems}, and a low magnetic ordering temperature not exceeding 26~K \cite{otrokov2019firstAFMMBT, zeugner2019tempMBT}.

One approach to mitigating the problem of bulk conductivity is the isovalent substitution of bismuth by antimony, leading to the formation of $\mathrm{MnSb_2Te_4}$ (MST). Crystals of $\mathrm{MnSb_2Te_4}$ are structural analogs of $\mathrm{MnBi_2Te_4}$ \cite{aliev2019mainrecipe,maksimov2023ramanspectrum,eremeev2021MSTreview}. This structural similarity is confirmed by X-ray diffraction studies as well as by investigations of solid solutions $\mathrm{MnBi_{0.5}Sb_{1.5}Te_4}$ \cite{abdullayev2021MBSTanalogue}.

The magnetic ground state of stoichiometric MST is an $A$-type antiferromagnetic order \cite{zang2022AFMtoFM,yan2019MBTMSTevol}. However, antisite defects, in particular $\mathrm{Mn_{Sb}}$, play a crucial role in determining the magnetic properties of MST, and their presence induces ferromagnetic ordering \cite{liu2021sitemixingMSTMBT,hu2021defectstuningMST}. Moreover, controlled manganese enrichment and the concomitant increase in the concentration of antisite defects can enhance the Curie temperature ($T_{\mathrm{C}}$) up to $45$--$50\,\mathrm{K}$ \cite{wimmer2021mn-richhightempMST}, thereby opening a route toward the realization of the QAHE at temperatures well above that of liquid helium.

Previous studies of the magnetic structure of MST using magnetic force microscopy (MFM) have focused on domain walls in the antiferromagnetic phase \cite{sass2020MBTMSTwalls}, the direct visualization of ferromagnetism in samples with $T_{\mathrm{C}} \approx 33\,\mathrm{K}$ \cite{ge2021directFMMST}, and the control of stripe and bubble domains in ferrimagnetic films with $T_{\mathrm{C}} \approx 27\,\mathrm{K}$ \cite{webb2024MSTdomains}. However, the nature of the domain structure in bulk MST single crystals with elevated $T_{\mathrm{C}}$ remains an open question, as an increased concentration of antisite defects may give rise to pronounced magnetic inhomogeneity. Furthermore, alternative magnetic imaging techniques, such as the Bitter decoration technique \cite{vinnikov2019decoration}, which eliminates the influence of a magnetic probe on the domain structure, have not yet been employed to study these systems.

In this work, we investigate the domain structure of a $\mathrm{MnSb_2Te_4}$ single crystal with a Curie temperature $T_{\mathrm{C}} \approx 45\,\mathrm{K}$, using the high-resolution Bitter decoration technique with magnetic nanoparticles. A domain structure characterized by two distinct spatial length scales is observed. The results are discussed in the context of an inhomogeneous distribution of $\mathrm{Mn_{Sb}}$ antisite defects, with a possible additional contribution from inversion symmetry breaking in the near-surface layer.

\section{Methods}
\label{methods}

A $\mathrm{MnSb_2Te_4}$ single crystal was obtained by cleaving from an ingot grown by the directional crystallization method \cite{aliev2019mainrecipe}. The sample had the form of a thin plate with dimensions of $\sim 3.5 \times 3.5 \times 0.16$ mm$^3$. Sample characterization was performed by X-ray diffraction and Raman scattering at room temperature \cite{maksimov2023ramanspectrum}.

Electrical contacts for transport measurements were fabricated using conductive graphite paste. The measurements were performed in a liquid-helium cryostat equipped with a superconducting solenoid. The external magnetic field was applied along the crystallographic $\mathbf{c}$ axis, i.e., perpendicular to the basal plane of the sample. The electrical resistance and Hall effect were measured in a standard four-probe geometry using an AC lock-in technique at a frequency of 20~Hz.

Prior to the decoration experiments, the surface of the single crystal was prepared by mechanical exfoliation using adhesive tape to remove surface contamination and oxide layers.

To visualize the domain structure in a bulk $\mathrm{MnSb_2Te_4}$ single crystal, the high-resolution Bitter decoration technique using iron magnetic nanoparticles was employed \cite{vinnikov2019decoration}. The experiments were carried out using a cryogenic insert without active sample temperature control. Compared to MFM, this technique offers two key advantages: it is less sensitive to surface quality and eliminates artifacts associated with the direct influence of a magnetic probe on the domain structure.

The initial parameters in the decoration chamber were: helium pressure $\sim 7 \times 10^{-2}$ Torr at a temperature of $4.3$ K. The nanoparticle deposition process was initiated by a rectangular current pulse with an amplitude of $12$ A and a duration of $2.2 \pm 0.3$ s applied to the evaporator. During deposition, the pressure in the chamber increased to $1.1 \times 10^{-1}$ Torr, and the temperature reached $\sim 6$ K.

A niobium (Nb) film with a thickness of $240\,\mathrm{nm}$ and a superconducting transition temperature $T_{\mathrm{sc}} \approx 9\,\mathrm{K}$ was used as a control sample.

The samples were cooled to the experimental working temperature in zero external magnetic field ($B_{\mathrm{ext}} \approx 0$, zero-field cooling, ZFC). After thermal equilibrium was established at $T = 4.3\,\mathrm{K}$, an external magnetic field of $50$~Oe, oriented perpendicular to its basal plane, was applied to the single crystal. The magnetic field was generated by a solenoid with a copper winding.

\section{Results and Discussion}
\label{results and discussion}

\begin{figure}[t]
    \centering

    \includegraphics[width=\columnwidth]{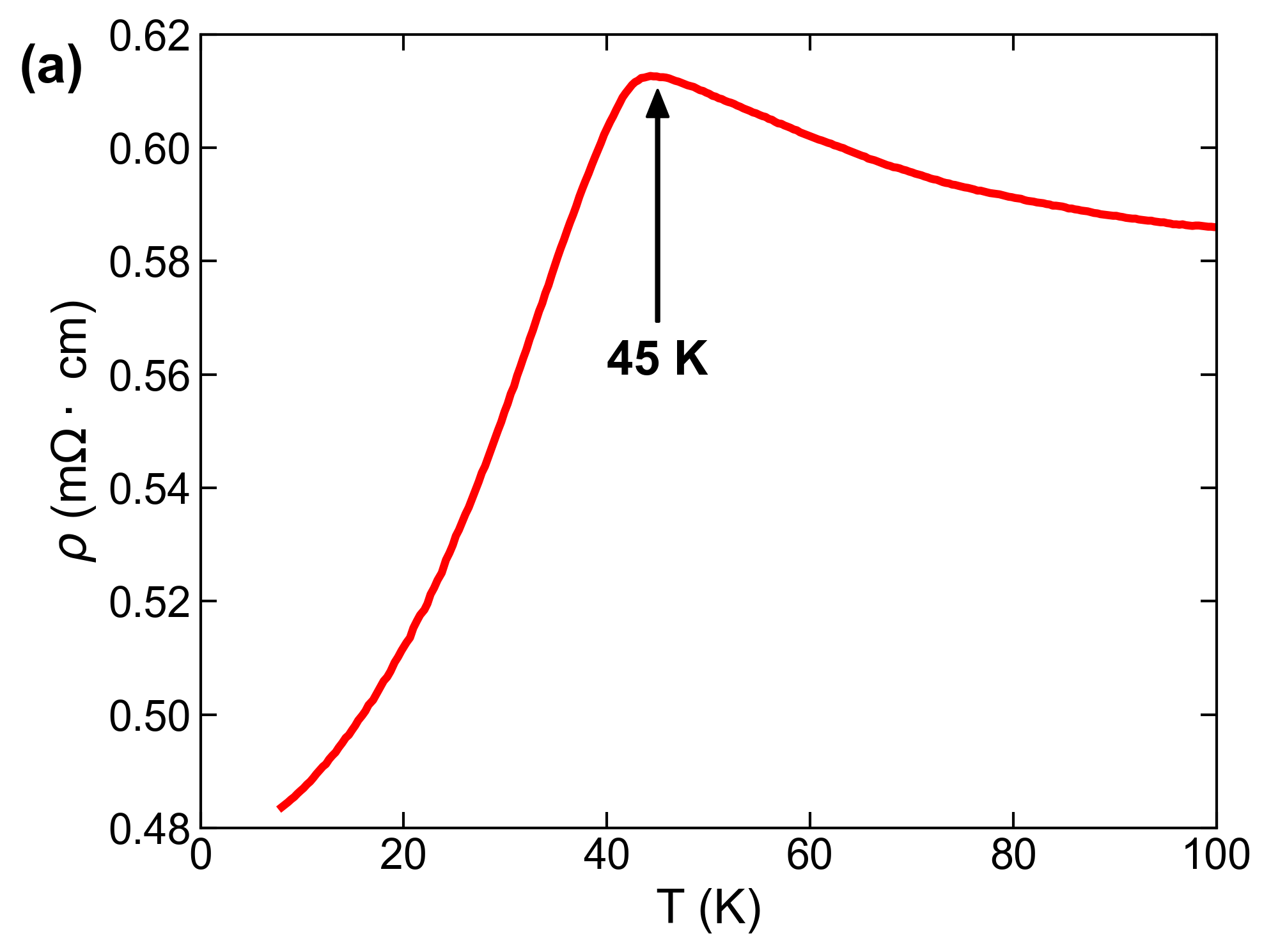}

    \vspace{2mm}

    \includegraphics[width=\columnwidth]{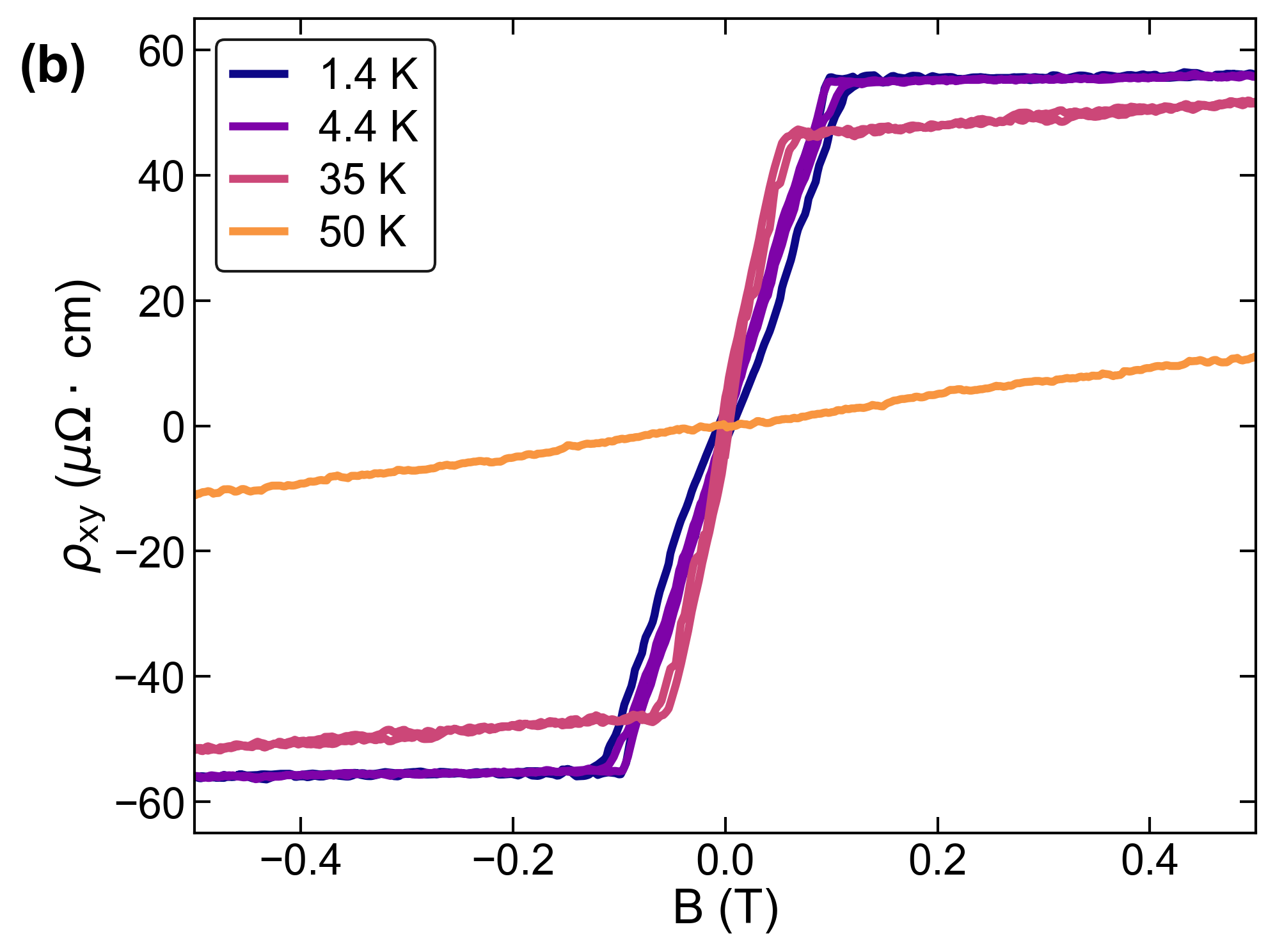}

    \caption{\justifying(Color Online) Magnetotransport properties of a $\mathrm{MnSb_2Te_4}$ single crystal. (a) Temperature dependence of the resistivity $\rho(T)$ measured in zero magnetic field. (b) Transverse (Hall) resistivity $\rho_{xy}$ as a function of the perpendicular magnetic field $B$, measured at different temperatures.}
    \label{fig:magnetic_data}
\end{figure}

The investigated $\mathrm{MnSb_2Te_4}$ single crystal exhibits ferromagnetic behavior. Figure \ref{fig:magnetic_data}shows the results of magnetotransport measurements. The temperature dependence of the resistivity $\rho(T)$ \cite{zverev2026magnetotransport} showed a maximum near $T_{\mathrm{C}} \approx 45\,\mathrm{K}$, which corresponds to the paramagnetic to ferromagnetic phase transition.

\begin{figure*}[t]
\centering

% ---------------- Row 1 ----------------
\begin{overpic}[width=0.32\textwidth]{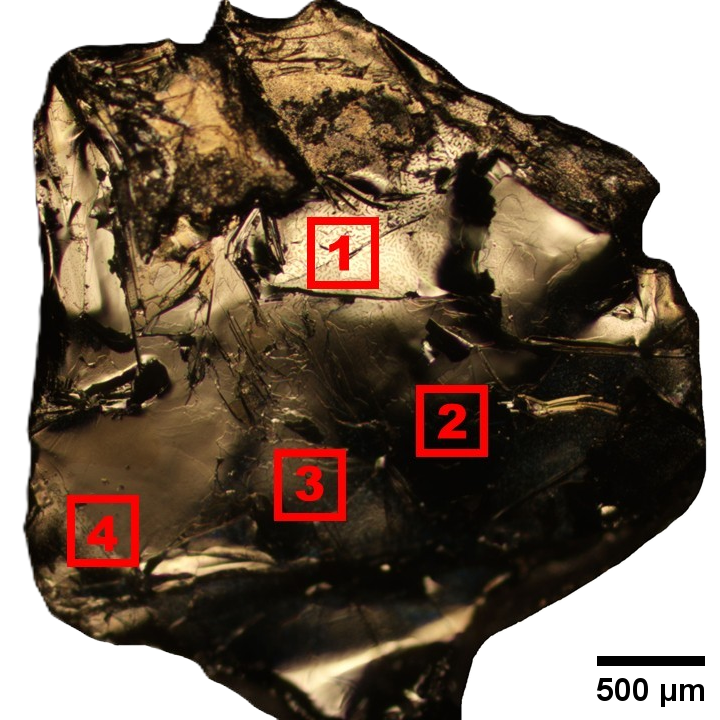}
  \put(2,92){\textbf{(a)}}
\end{overpic}\hfill
\begin{overpic}[width=0.32\textwidth]{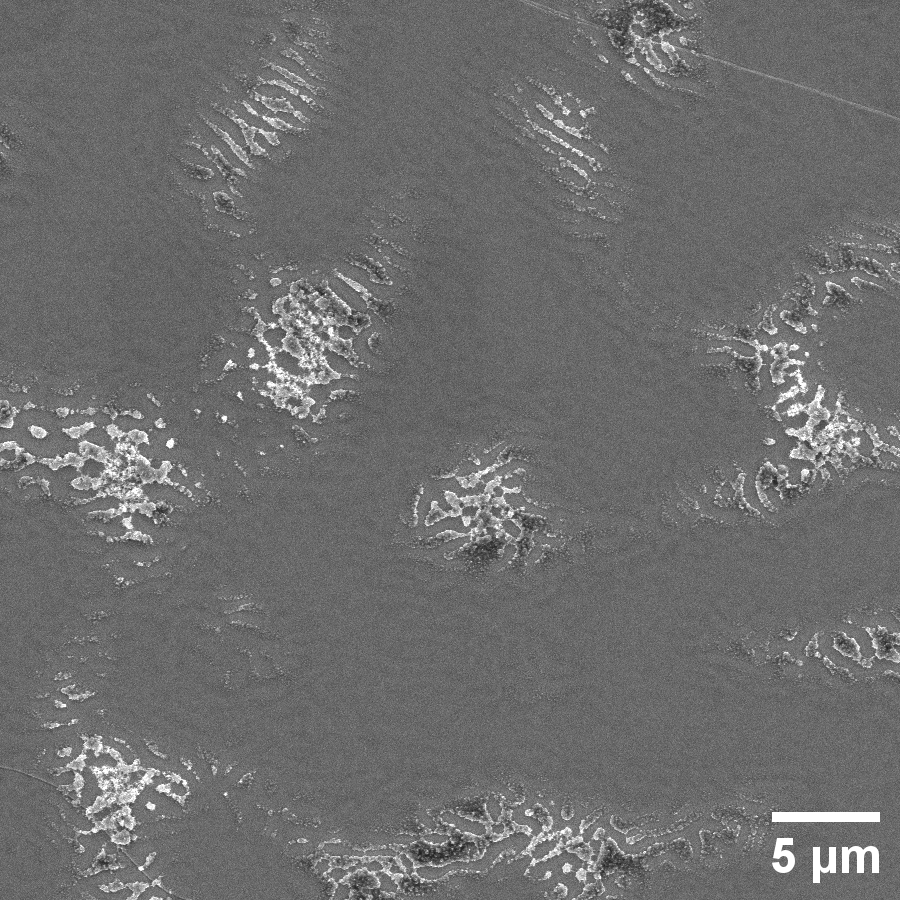}
  \put(2,92){\textbf{\color{white}(b)}}
\end{overpic}\hfill
\begin{overpic}[width=0.32\textwidth]{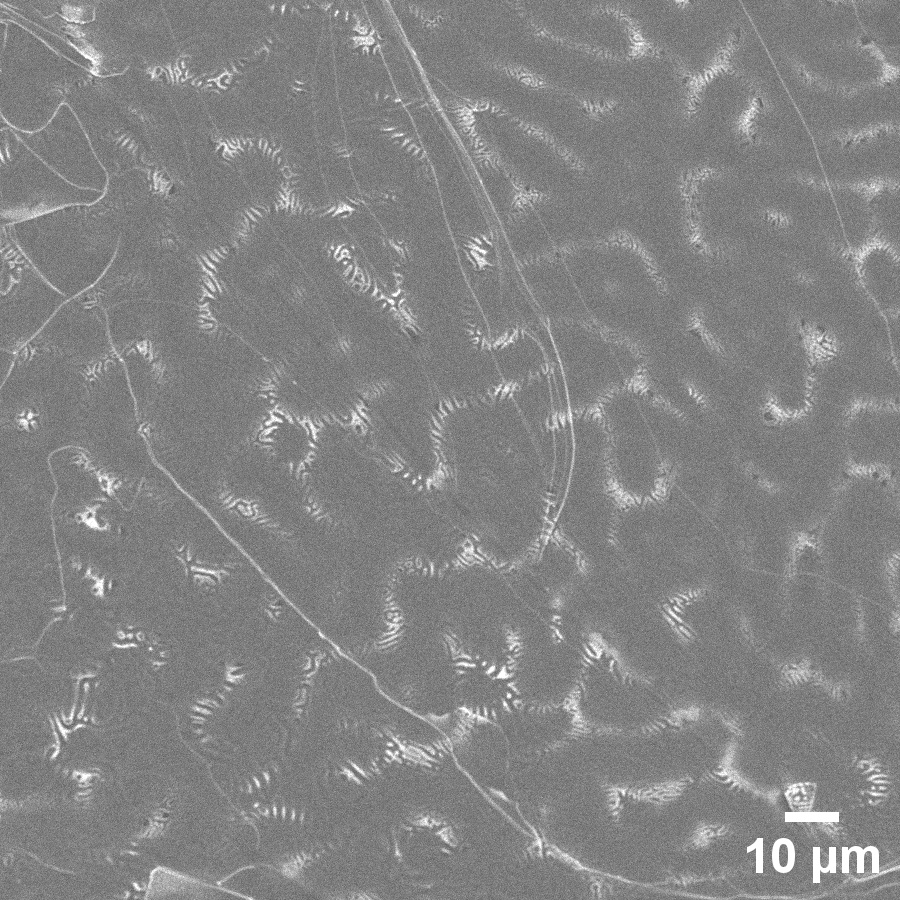}
  \put(2,92){\textbf{\color{white}(c)}}
\end{overpic}

\vspace{2mm}

% ---------------- Row 2 ----------------
\begin{overpic}[width=0.32\textwidth]{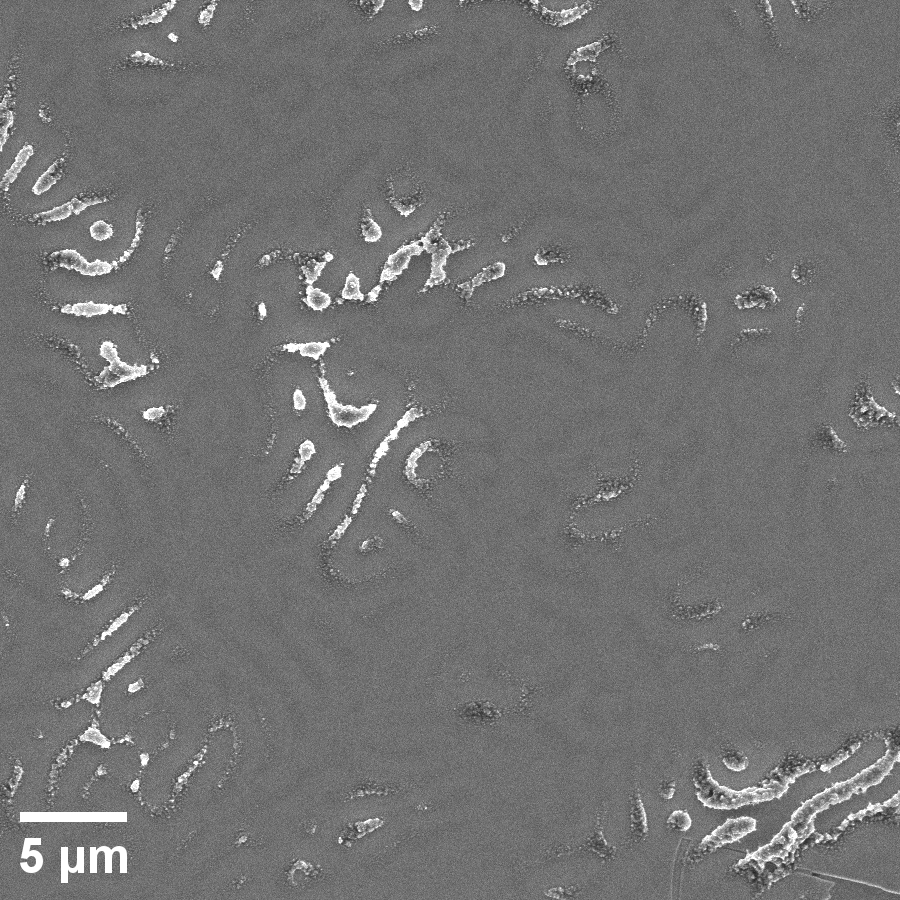}
  \put(2,92){\textbf{\color{white}(d)}}
\end{overpic}\hfill
\begin{overpic}[width=0.32\textwidth]{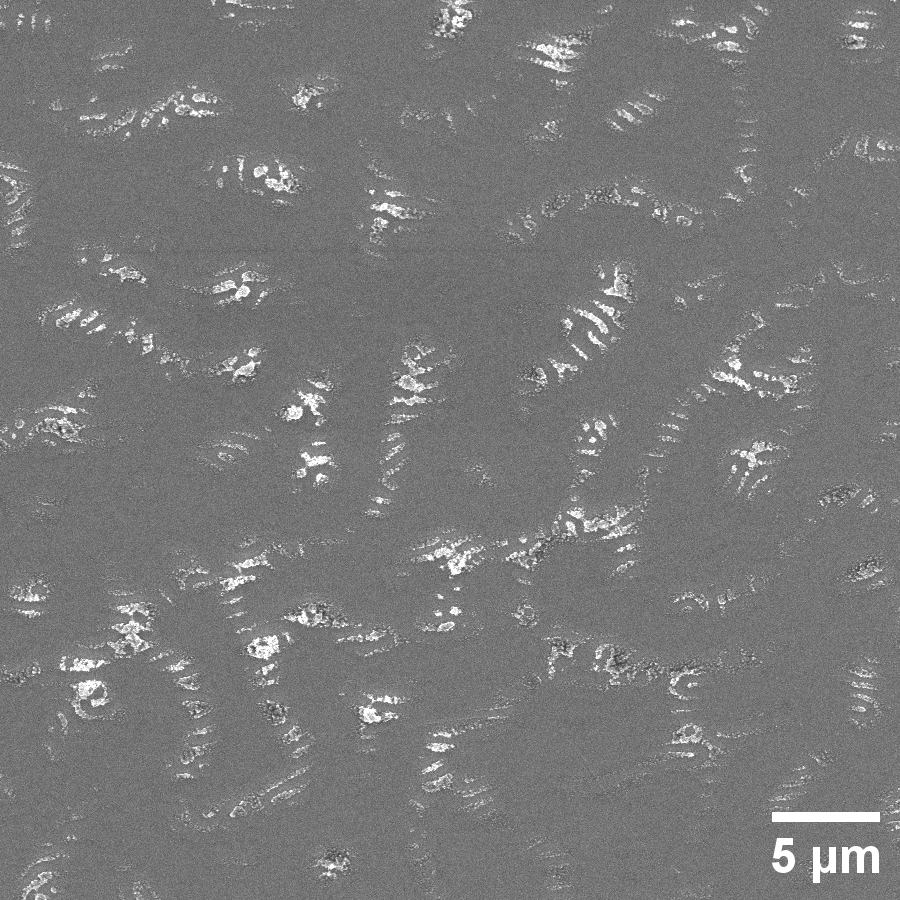}
  \put(2,92){\textbf{\color{white}(e)}}
\end{overpic}\hfill
\begin{overpic}[width=0.32\textwidth]{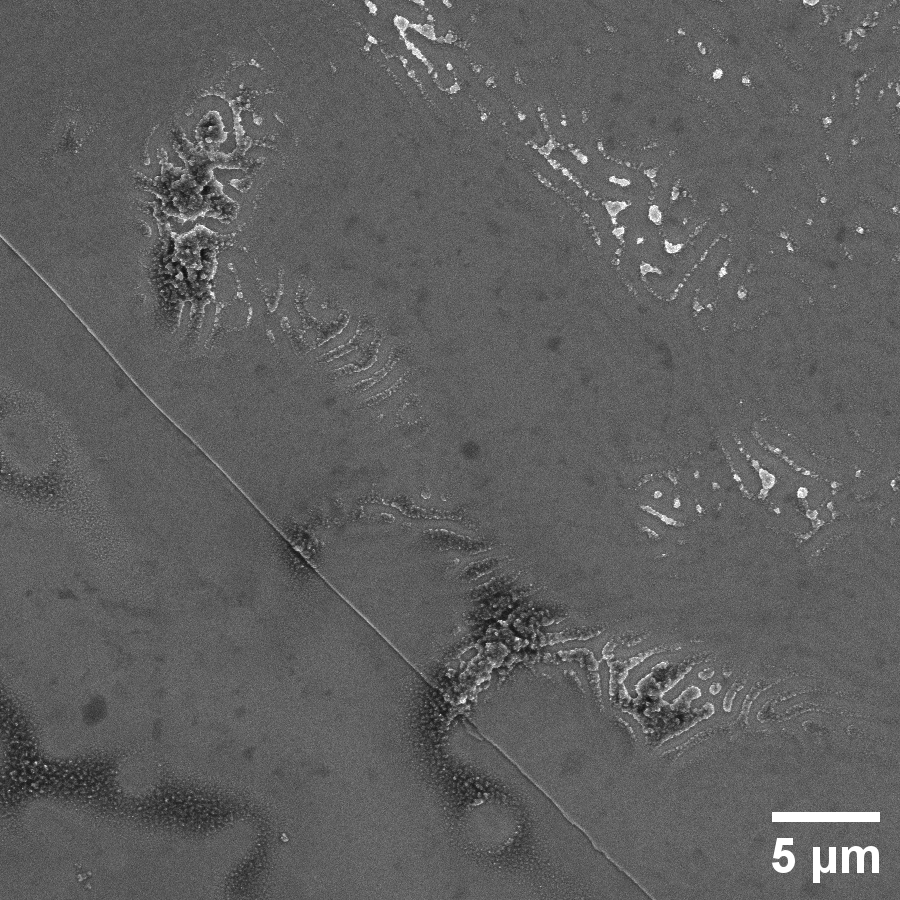}
  \put(2,92){\textbf{\color{white}(f)}}
\end{overpic}

\caption{\justifying(Color Online) Micrographs of the surface and decoration patterns of a $\mathrm{MnSb_2Te_4}$ single crystal. (a) Overview of the crystal surface obtained by optical microscopy. (b)--(f) Images of the magnetic flux structure in different regions of the sample obtained by decoration with magnetic nanoparticles and visualized by scanning electron microscopy: (b) region~1; (c) region~2; (d,e) region~3; (f) region~4.}
\label{fig:subdomains}
\end{figure*}

The presence of spontaneous magnetization in the ferromagnetic phase is confirmed by the observation of the Hall resistivity $\rho_{xy}$ (Fig. \ref{fig:magnetic_data}). The $\rho_{xy}(B)$ dependencies exhibit hysteretic behavior with a small coercive field $H_{\mathrm{c}} \sim 100$~Oe, which is characteristic of soft magnetic materials.

The morphology of the domain structure was investigated using the high-resolution Bitter decoration technique. In the control sample Nb film, zero-field cooling (ZFC) revealed a dendritic magnetic flux penetration pattern typical of type-II superconductors \cite{duran1995observation}.

Figure~\ref{fig:subdomains} presents the results of magnetic flux visualization in $\mathrm{MnSb_2Te_4}$ for different regions of the single crystal with a thickness of $160 \pm 20\,\mu\mathrm{m}$. In certain surface regions (areas~1--4), a two-scale domain structure is observed, while the majority of the surface exhibits a typical stripe domain pattern with isolated bubble inclusions, characteristic of ferromagnets with strong uniaxial anisotropy~\cite{hubert1998domains}. The two-scale domain structure is spatially localized and covers less than $10\%$ of the investigated surface. However, it is reproducible in repeated experiments after exfoliation of the surface layer (typically on the order of $\sim 10\,\mu\mathrm{m}$), indicating the robustness of the observed effect.

The observed decoration patterns reflect the distribution of magnetic flux above the sample surface, which is confirmed by several independent experimental observations. In particular, domain stripes continuously cross natural crystal surface steps with height variations of tens of nanometers (Fig.~\ref{fig:subdomains}(c)), and qualitatively similar decoration patterns are reproducibly obtained in independent experiments carried out under identical conditions.

The contrast in decoration images is governed by the spatial distribution of the stray magnetic field above the sample surface. Iron nanoparticles, being in a superparamagnetic or ferromagnetic state, become magnetized under the action of an external magnetic field and are preferentially deposited in regions with the maximum normal component of the stray field \cite{sakurai1992decoration}. As a result, the local particle density is proportional to the magnetic flux density, providing a direct correspondence between the decoration pattern and the domain structure.

Differences in contrast between images acquired by optical microscopy (OM) and scanning electron microscopy (SEM) arise from the distinct image-formation mechanisms of these techniques. In OM, regions with a high density of deposited particles appear dark, whereas in SEM the same regions appear brighter.

The observed features of the domain structure exhibit signatures of self-similarity: large-scale domains with a characteristic period on the order of tens of micrometers contain embedded sub-domain patterns.

\begin{figure}[t]
\centering

% ---------------- Row 1 ----------------
\begin{overpic}[width=0.48\columnwidth]{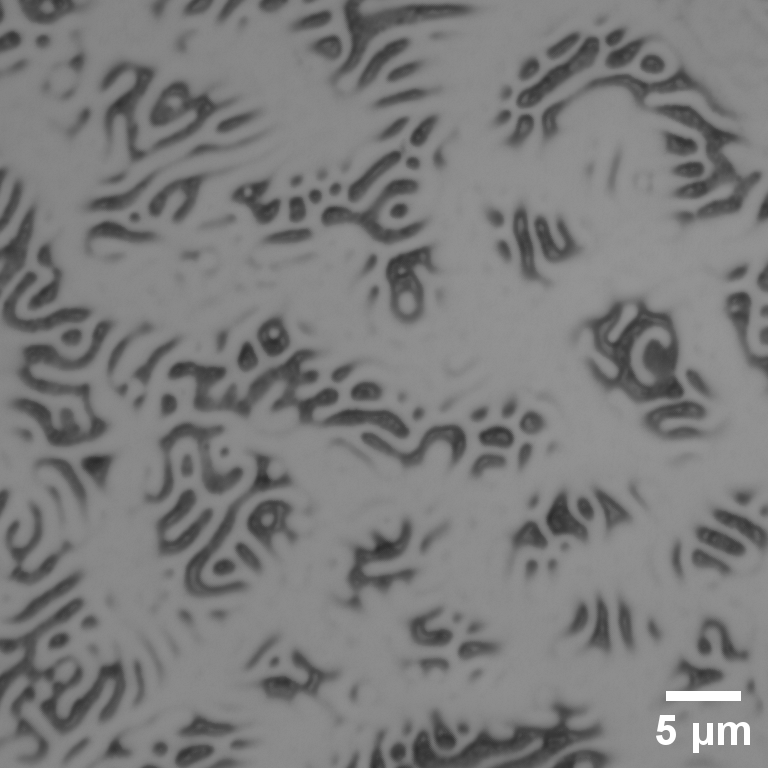}
  \put(2,92){\textbf{\color{white}(a)}}
\end{overpic}\hfill
\begin{overpic}[width=0.48\columnwidth]{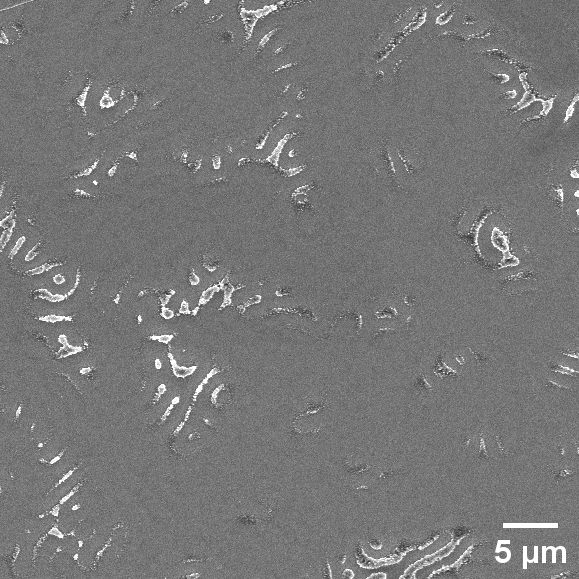}
  \put(2,92){\textbf{\color{white}(b)}}
\end{overpic}

\vspace{2mm}

% ---------------- Row 2 ----------------
\begin{overpic}[width=0.48\columnwidth]{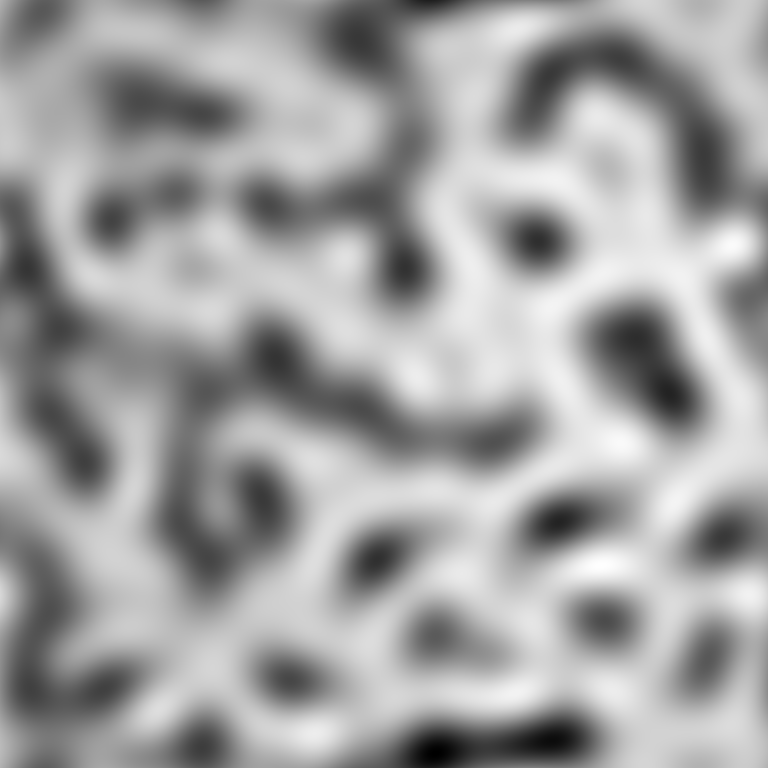}
  \put(2,92){\textbf{\color{white}(c)}}
\end{overpic}\hfill
\begin{overpic}[width=0.48\columnwidth]{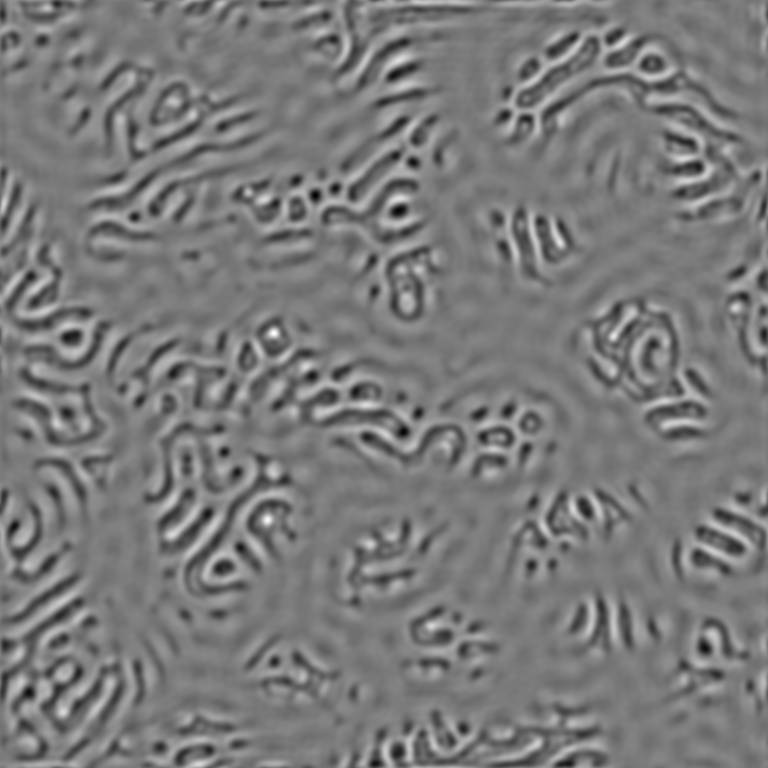}
  \put(2,92){\textbf{\color{white}(d)}}
\end{overpic}

\vspace{2mm}

% ---------------- Row 3 ----------------
\begin{overpic}[width=0.48\columnwidth]{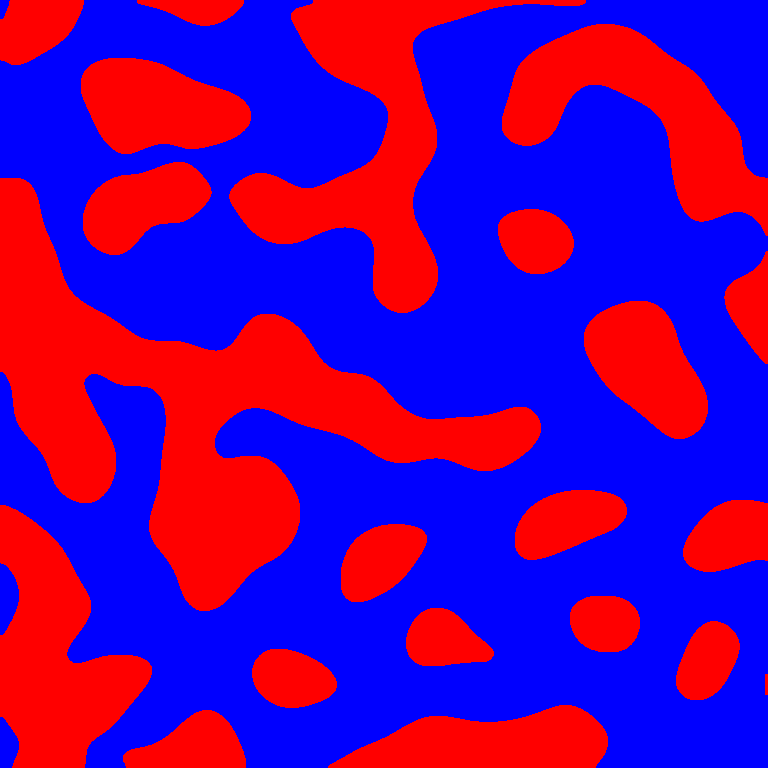}
  \put(2,92){\textbf{\color{white}(e)}}
\end{overpic}\hfill
\begin{overpic}[width=0.48\columnwidth]{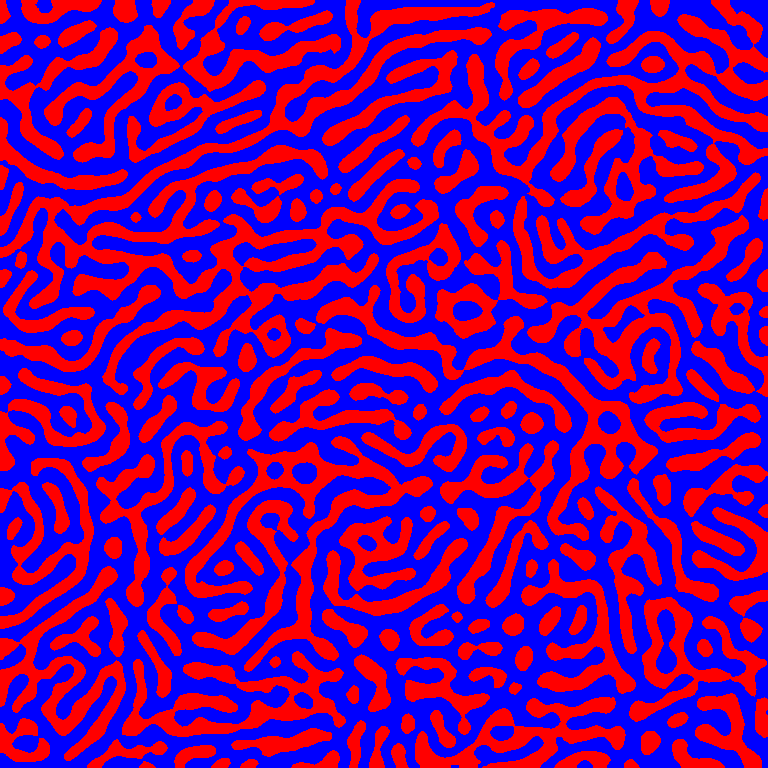}
  \put(2,92){\textbf{\color{white}(f)}}
\end{overpic}

\caption{\justifying(Color Online) Comparative analysis of the domain structure in $\mathrm{MnSb_2Te_4}$. (a,b) Domain morphology obtained after decoration with magnetic nanoparticles, visualized by optical microscopy (a) and scanning electron microscopy (b). (c,d) Spatial frequency analysis highlighting the large-scale (c) and small-scale (d) components of the domain structure. (e,f) Binarized images corresponding to the large-scale (e) and small-scale (f) domain textures.}

\label{fig:FFT_analysis}
\end{figure}

For a quantitative characterization of the domain structure, Fourier analysis of the optical images was performed (Fig.~\ref{fig:FFT_analysis}). The resulting spatial spectra reveal distinct characteristic frequencies corresponding to two spatial scales of the magnetic texture.

After binarization of the images (Fig.~\ref{fig:FFT_analysis}(e),~\ref{fig:FFT_analysis}(f)), the average periods of the domain structure were determined to be $\lambda_1 = 14.2\,\mu\mathrm{m}$ with $\Delta \lambda_1 = 1.1\,\mu\mathrm{m}$ for the large-scale component and $\lambda_2 = 1.82\,\mu\mathrm{m}$ with $\Delta \lambda_2 = 0.24\,\mu\mathrm{m}$ for the small-scale modulation. The ratio of the characteristic periods is approximately $7.8$, indicating the formation of a hierarchical domain structure with two clearly separated spatial length scales.

A comparison of images obtained by optical microscopy and scanning electron microscopy (Fig.~\ref{fig:FFT_analysis}(a),~\ref{fig:FFT_analysis}(b)) demonstrates good agreement in the overall domain morphology. Minor differences in the apparent domain width can be attributed to optical diffraction broadening.

The observed Curie temperature, $T_{\mathrm{C}} \approx 45\,\mathrm{K}$, is in good agreement with the data of~\cite{wimmer2021mn-richhightempMST}, where similar temperatures were achieved in Mn-enriched $\mathrm{MnSb_2Te_4}$ samples. This consistency confirms that the controlled introduction of $\mathrm{Mn_{Sb}}$ antisite defects provides an effective route to enhancing the Curie temperature in this class of materials.

Indirect evidence for stratification of magnetic subsystems has previously been reported based on features of the anomalous Hall effect in doped vanadium crystals \cite{fijalkowski2020bulksurfaceAHE}. However, direct visualization of a sub-domain structure has so far been achieved only in the van der Waals magnet $\mathrm{CrBr_3}$ using MFM \cite{grebenchuk2025subdomains}. In that case, the effect was relatively weak, likely due to the limited sensitivity of the technique. In the present study, optical microscopy after decoration enables clear identification of a two-scale domain structure.

The physical origin of magnetic stratification remains unclear. In $\mathrm{CrBr_3}$, it has been associated with the formation of a magnetic barrier at stacking faults, a concept that was generalized to van der Waals crystals as a whole \cite{grebenchuk2025subdomains}. However, for compounds of the $\mathrm{Mn(Sb,Bi)_2Te_4}$ family, experimental evidence for such stacking faults is lacking \cite{liu2021sitemixingMSTMBT,hu2021defectstuningMST}.

We propose that the observed separation of magnetic subsystems into surface and bulk components may be related to the formation of a magnetic barrier, analogous in its effect to the stacking-fault-induced barrier discussed for $\mathrm{CrBr_3}$ \cite{grebenchuk2025subdomains}. In the case of $\mathrm{MnSb_2Te_4}$, such a barrier is likely governed by a different microscopic mechanism and may originate from an inhomogeneous distribution of $\mathrm{Mn_{Sb}}$ antisite defects. An additional contribution may arise from symmetry lowering in the near-surface layer, resulting from inversion symmetry breaking and the formation of point defects during exfoliation \cite{sass2020robust,shikin2020exfoliation}. These effects can lead to modifications of exchange and dipolar interactions, as well as of the magnetic anisotropy \cite{rusinov2021surface}.

As a result, two magnetically weakly coupled subsystems with different effective magnetic layer thicknesses are formed. The observed self-similarity of the domain structure across both spatial scales is consistent with the fact that domain morphology is governed by the balance of exchange, dipolar, magnetostatic interactions, and magnetic anisotropy. These interactions are qualitatively similar in the bulk and near the surface, but scale quantitatively with the effective layer thickness. Consequently, the revealed sub-domain morphology (Fig.~\ref{fig:FFT_analysis}) can be interpreted as a superposition of two self-similar domain structures formed in weakly coupled surface and bulk magnetic layers. The precise microscopic mechanism underlying such stratification requires further experimental and theoretical investigation.

\section{Conclusion}
\label{conclusion}

In this work, a sub-domain magnetic structure characterized by two distinct spatial scales was discovered in a $\mathrm{MnSb_2Te_4}$ single crystal with a Curie temperature of $\sim 45\,\mathrm{K}$ using the high-resolution Bitter decoration technique. The observed qualitative difference in domain morphology compared to previously studied samples with lower Curie temperatures suggests a possible link between the ferromagnetic ordering temperature and the characteristics of the domain structure.

The two-scale domain structure is interpreted as resulting from the superposition of two magnetically weakly coupled subsystems with different effective magnetic layer thicknesses and characteristic domain sizes. A plausible mechanism involves the formation of a magnetic barrier associated with an inhomogeneous distribution of $\mathrm{Mn_{Sb}}$ antisite defects, with an additional contribution from inversion symmetry breaking in the near-surface layer. It should be emphasized that this interpretation is qualitative and requires further experimental and theoretical verification.

A promising direction for future research is a systematic investigation of the domain structure across a series of samples with different Curie temperatures, which would enable testing a possible correlation between magnetic ordering parameters and domain morphology. Complementary micromagnetic modeling aimed at elucidating the role of defects and surface effects in van der Waals magnetic topological materials could provide further insight into the mechanisms responsible for the observed structure.

\section*{Acknowledgments}

The authors are grateful to S.~V.~Egorov and A.~S.~Astakhantseva for conducting the SEM studies. We also thank the team of the laboratory of Superconductivity at the Institute of Solid State Physics, Russian Academy of Sciences, and in particular V.~V.~Ryazanov, N.~A.~Tulina, and A.~S.~Ionin for fruitful discussions.

\section*{Funding}

This work was carried out within the framework of the state assignment of the Yu.\,A.~Osipyan Institute of Solid State Physics of the Russian Academy of Sciences and was also supported by the state budget of the Institute of Physics of the Ministry of Science and Education of Azerbaijan.

\section*{Conflict of interest}

The authors declare no conflict of interest.

\bibliography{bibtex}

\end{document}